\documentclass[aps,prd,onecolumn,eqsecnum,amsmath,nofootinbib,preprintnumbers]{revtex4}%
\newcounter{example}[section]

\usepackage{amsmath}
\usepackage{color,graphicx,float,subfigure}
\usepackage{amsfonts,amssymb,theorem,mathrsfs,times}
\usepackage{bm}
\usepackage{amsmath}
{\theorembodyfont{\upshape}
	}
{\theorembodyfont{\upshape}
	}
{\theorembodyfont{\upshape}
	}
{\theorembodyfont{\upshape}
	}
{\theorembodyfont{\upshape}
	}
{\theorembodyfont{\upshape}
	}

\newcommand{\dalm}{\kern1pt\vbox{\hrule height 0.9pt\hbox{\vrule width
			0.9pt\hskip 2.5pt\vbox{\vskip 5.5pt}\hskip 3pt\vrule width
			0.3pt}\hrule height 0.3pt}\kern1pt}

\begin{document}
	\title{Bounds on the photon sphere radius for spherically symmetric black holes in $\boldsymbol{n}$-dimensional Einstein gravity}
	
	%
	\author{Yong Song\footnote{e-mail
			address: syong@cdut.edu.cn (corresponding author)}
	}
	\author{Jiaqi Fu\footnote{e-mail
			address: 1491713073@qq.com}}
	\author{Yiting Cen\footnote{e-mail
			address: 2199882193@qq.com}}

	
	\affiliation{
		College of Physics\\
		Chengdu University of Technology, Chengdu, Sichuan 610059,
		China}
	

	\date{\today}
	
\begin{abstract}
The photon sphere, a hypersurface of null circular geodesics, plays a fundamental role in characterizing black hole spacetimes, influencing phenomena such as black hole shadows, gravitational lensing, and quasinormal modes. In this work, we derive both upper and lower bounds on the photon sphere radius for static, spherically symmetric, asymptotically flat black holes within $n$-dimensional Einstein gravity ($n\ge 4$), assuming an anisotropic matter field satisfying the weak energy condition and a non‑positive trace of the energy‑momentum tensor. For the upper bound, we obtain $r_\gamma\le [(n-1)M]^{\frac{1}{n-3}}$, where $M$ is the ADM mass. In the four-dimensional case ($n=4$), this reduces to $r_\gamma\le 3M$, in agreement with previous results. For the lower bound, under the additional assumption that $|r^{n-1}p_r(r)|$ is monotonically decreasing, we prove $r_\gamma\ge (\frac{n-1}{2})^{1/(n-3)}r_H$, where $r_H$ is the radius of the outer event horizon; for $n=4$ this gives $r_\gamma\ge \frac{3}{2}r_H$, also consistent with previous four‑dimensional result. These results provide dimension‑dependent geometric constraints that generalize well‑known four‑dimensional bounds to a specific class of higher‑dimensional black holes (described by a Tangherlini‑type metric) and deepen our understanding of spacetime structure in higher‑dimensional gravitational theories.
\end{abstract}
	

\maketitle

	
\section{Introduction}
Black holes, as fundamental predictions of general relativity and key astrophysical objects, continue to be a central focus of theoretical and observational research. The direct image of a black hole shadow, captured by the Event Horizon Telescope (EHT)~\cite{EventHorizonTelescope:2019dse,EventHorizonTelescope:2022wkp}, has spectacularly confirmed the existence of these compact objects and our understanding of strong-field gravity. Among the characteristic features of black hole spacetimes, the photon sphere---a hypersurface on which massless particles can orbit the black hole on unstable circular null geodesics---plays a crucial role. It determines the boundary of the black hole shadow~\cite{Perlick:2010zh,Falcke:1999pj}, is intimately connected to the characteristic quasinormal modes of black holes~\cite{Cardoso:2008bp,Berti:2009kk}, and constrains the spatial extent of matter fields (hair) outside the horizon~\cite{Nunez:1996xv,Hod:2011aa}.
	
For the four-dimensional Schwarzschild black hole, the photon sphere is located at $r_\gamma=3M$. For more general static, spherically symmetric, asymptotically flat black holes surrounded by matter, the photon sphere radius satisfies the upper bound $r_\gamma\le 3M$ provided the matter obeys the weak energy condition (or null energy condition) and its energy‑momentum tensor has a non‑positive trace~\cite{Hod:2013jhd,Yang:2019zcn}. This bound is saturated by the vacuum (bald) Schwarzschild black hole. Subsequently, a lower bound $r_\gamma\ge\frac{3}{2}r_H$ was established under the additional condition that $|r^{3}p_r(r)|$ decreases monotonically~\cite{Hod:2020pim,Yang:2019zcn}. Further studies have explored related upper and lower bounds for photon spheres of compact stars~\cite{Lu:2019zxb,Peng:2020shc,Liu:2024odv,Hod:2023jmx,Song:2025fgq}. These bounds are not only of theoretical interest but also provide insights into the possible size of black hole shadows and the structure of hairy black holes.
	
Given the significant interest in higher-dimensional theories of gravity---motivated by string theory, brane-world models, and the gauge/gravity duality---a natural question arises: do similar bounds exist for the photon sphere in higher-dimensional black hole spacetimes? Understanding how spacetime dimensionality affects fundamental structures like the photon sphere is essential for probing the geometry of higher-dimensional gravity.
	
In this work, we fill this gap by deriving both an upper and a lower bound for the photon sphere radius in static, spherically symmetric, asymptotically flat black holes of $n$-dimensional Einstein gravity ($n\ge 4$). Our analysis is performed within a model‑independent framework that assumes a Tangherlini‑type metric structure and an anisotropic matter fluid satisfying the weak energy condition and a non‑positive trace of the energy‑momentum tensor. For the upper bound, we obtain
\begin{align}
r_\gamma\le [(n-1)M]^{\frac{1}{n-3}}\;,
\end{align}
which for $n=4$ reduces to $r_\gamma\le 3M$. For the lower bound, we impose one further physically motivated condition---the monotonic decrease of $|r^{n-1}p_r(r)|$---and prove
\begin{align}
r_\gamma\ge \bigg(\frac{n-1}{2}\bigg)^{1/(n-3)}r_H\;,
\end{align}
which for $n=4$ becomes $r_\gamma\ge \frac{3}{2}r_H$. These results provide a complete set of dimension‑dependent geometric constraints that generalise the well‑known four‑dimensional bounds to a specific class of higher‑dimensional black holes.

The paper is organized as follows: In Sec.~\ref{section2}, we describe the general higher-dimensional black hole spacetime and the corresponding Einstein field equations for an anisotropic fluid. In Sec.~\ref{section3}, we derive the photon sphere condition, prove its existence, and establish the upper bound under appropriate energy conditions. In Sec.~\ref{section4}, we derive the lower bound, employing the monotonicity property of the radial pressure function. We conclude with a discussion of the results, their limitations, and possible future extensions in Sec.~\ref{conclusion}. In App.~\ref{Appendix}, we provide a rigorous proof of an inequality needed in the lower bound analysis. Throughout, we use natural units with $G=c=1$.


\section{Description of the system in the higher-dimensional black hole}\label{section2}
We begin by recalling the general metric ansatz for a spacetime $(\mathcal{M}, g)$ that possesses the symmetry of a codimension-2 maximally symmetric space, which reads~\cite{Cao:2019vlu}
\begin{equation}
	\label{metric}
	g=h_{AB}(y)dy^Ady^B + r^2(y)\gamma_{ij}(z)dz^idz^j\, ,
\end{equation}
where $A=1,2$, and $i=1,\cdots \, ,n-2$, and $\gamma_{ij}dz^idz^j$ is the metric of the codimension-2 maximally symmetric space $(\mathcal{K},\gamma)$ with a sectional curvature $k=0\, ,\pm 1$. The two dimensional part of $(\mathcal{M},g)$ with coordinates $\{y^A\}$ has a Lorentz signature and can be denoted by $(M, h)$. 

The connection coefficients of the metric (\ref{metric}) are
\begin{align}
&\Gamma^{A}{}_{BC}={}^{(2)}\Gamma^{A}{}_{BC}(y),\quad\Gamma^i{}_{jk}=\hat{\Gamma}^{i}{}_{jk}(z)\;,\\
&\Gamma^{A}{}_{ij}=-r(D^Ar)\gamma_{ij},\quad\Gamma^i{}_{jA}=\frac{D_Ar}{r}\delta^i{}_j\;,
\end{align}
where ${}^{(2)}\Gamma$ denotes the connection coefficients computed from the two-dimensional Lorentzian metric $(M, h)$, $D_A$ is the covariant derivative compatible with $h_{AB}$ on $(M, h)$, and $\hat{\Gamma}^{i}{}_{jk}$ are connection coefficients of the codimension-2 maximally symmetric space $(\mathcal{K},\gamma)$.

The components of the Riemann tensor are
\begin{align}
&R_{ABCD}={}^{(2)}R_{ABCD}\;,\\
&R_{AiBj}=-r(D_AD_Br)\gamma_{ij}\;,\\
&R_{ijkl}=r^2(1-D_ArD^Ar)(\gamma_{ik}\gamma_{jl}-\gamma_{il}\gamma_{jk})\;,
\end{align}
and the components of the Ricci tensor are
\begin{align}
&R_{AB}={}^{(2)}R_{AB}-(n-2)\frac{D_AD_Br}{r}\;,\\
&R_{ij}=[-r(D^AD_Ar)+(n-3)(1-D_ArD^Ar)]\gamma_{ij}\;,
\end{align}
while the Ricci scalar is
\begin{align}
R={}^{(2)}R-2(n-2)\frac{D^AD_Ar}{r}+\frac{(n-2)(n-3)}{r^2}(1-D_ArD^Ar)\;.
\end{align}
Consequently, the Einstein tensor can be expressed as
\begin{align}
\label{GAB}
&G^A{}_{B}=-\frac{(n-2)}{r}D^AD_Br-\frac{1}{2}\bigg[\frac{(n-2)(n-3)}{r^2}(1-D_CrD^Cr)-\frac{2(n-2)}{r}D_CD^Cr\bigg]h^A{}_{B}\;,\\
&G^i{}_j=\bigg\{-\frac{1}{2}{}^{(2)}R+(n-3)\frac{D_AD^Ar}{r}+\bigg[(n-3)-\frac{(n-2)(n-3)}{2}\bigg]\frac{1-D_ArD^Ar}{r^2}\bigg\}r^2\delta^i{}_j\;,\\
&G^A{}_i=0\;.
\end{align}
where we have used the fact that
\begin{align}
^{(2)}R_{AB}-\frac{1}{2}\ ^{(2)}Rh_{AB}=0\;,
\end{align}
for two dimensional spacetime.

In this work, we focus on static, spherically symmetric, and asymptotically flat higher-dimensional black hole systems within Einstein gravity. The $n$-dimensional Einstein-Hilbert action with matter fields is given by
\begin{align}
\label{S}
S=\int d^nx\sqrt{-g}\bigg(\frac{R}{16\pi}+\mathcal{L}_M\bigg) \;,
\end{align}
where $g$ is the determinant of the metric tensor, and $\mathcal{L}_M$ denotes the Lagrangian density of matter.

The metric in Ref.~\cite{Hod:2013jhd} can be easily generalized to a $n$-dimensional black hole spacetime and written as~\cite{Lee:2025xdm}
\begin{align}
\label{metrics}
ds^2=-e^{-2\delta(r)}\mu(r)dt^2+\mu(r)^{-1}dr^2+r^2d\Omega^2_{n-2}\;,
\end{align}
where the metric functions $\delta(r)$ and $ \mu(r)$ depend only on the areal coordinate $r$, and
\begin{align}
d\Omega^2_{n-2}=d\theta_1^2+\sin^2\theta_1d\theta_2^2+\cdots+\sin^2\theta_1\cdots \sin^2\theta_{n-3}d\theta_{n-2}^2\;,
\end{align}
represents the line element of the unit $(n-2)$-sphere.

The regularity of the event horizon at $r=r_H$ imposes the boundary conditions
\begin{align}
	\label{rH}
	\mu(r_H)=0\quad\mathrm{with}\quad \mu'(r_H)\geq 0\;.
\end{align}
Asymptotic flatness requires that as $r\rightarrow \infty$,
\begin{align}
	\label{infinity}
	\mu(r\rightarrow \infty)\rightarrow1\quad\mathrm{and}\quad \delta(r\rightarrow \infty)\rightarrow 0\;.
\end{align}
Here, we do not assume $\delta(r)=0$, so our results also apply to hairy black-hole configurations~\cite{Volkov:1998cc,Volkov:2016ehx}.

The matter field is modelled as an anisotropic fluid, whose energy‑momentum tensor is diagonal in the static orthonormal frame:
\begin{align}
T^\mu{}_{\nu}=\mathrm{diag}(-\rho(r),\,
p_r(r),\,
\underbrace{p_t(r),\,p_t(r),\,\ldots,\,p_t(r)}_{n-2\ \text{terms}})\;,
\end{align}
where $\rho$ is the energy density, $p_r$ is the radial pressure, and $p_t$ is the tangential pressure. From the Einstein field equations $G^{\mu}{}_{\nu}=8\pi T^{\mu}{}_{\nu}$ and using Eq.~(\ref{GAB}), we obtain
\begin{align}
\label{mu1}
&\mu'=\frac{(n-3)(1-\mu)}{r}-\frac{16\pi r\rho}{(n-2)}\;,\\
\label{delta1}
&\delta'=-\frac{8\pi r(\rho+p_r)}{(n-2)\mu}\;,
\end{align}
where the prime denotes differentiation with respect to $r$. Substituting Eq.~(\ref{rH}) into Eq.~(\ref{mu1}) and (\ref{delta1}) yields
\begin{align}
	\label{rhorH}
	\rho(r_H)\leq \frac{(n-3)(n-2)}{16\pi r_H^2}\;,\quad p_r(r_H)=-\rho(r_H)\;.
\end{align}
The mass $m(r)$ enclosed within a sphere of radius $r$ can be written as
\begin{align}
\label{m}
m(r)=\frac{1}{2}r_H^{n-3}+\frac{8\pi}{n-2}\int_{r_H}^{r}x^{n-2}\rho(x)dx\;,
\end{align}
where $\frac{1}{2}r_H^{n-3}$ is the  horizon mass $m(r_H)$. From Eqs.~(\ref{mu1}) and (\ref{m}), the relation between $\mu$ and $m(r)$ in $n$-dimensional spacetime can be expressed as
\begin{align}
	\label{mu}
	\mu(r)=1-\frac{2m(r)}{r^{n-3}}\;.
\end{align}
The condition for a finite mass configuration characteristics implies
\begin{align}
	\label{rho}
	\lim\limits_{r \to \infty}r^{n-1}\rho(r)=0\;.
\end{align}
Substituting Eqs.~(\ref{mu1}) and (\ref{delta1}) into the energy-momentum tensor conservation equation
\begin{align}
	\label{Tab}
	T^{\mu}_{r;\mu}=0\;,
\end{align}
yields
\begin{align}
	\label{pr1}
	p_r'=\frac{2T+(n-1)\rho-(n+1)p_r}{2r}-\bigg(\frac{n-3}{2r}+\frac{8\pi r p_r}{n-2}\bigg)\frac{\rho+p_r}{\mu}\;,
\end{align}
where
\begin{align}
\label{T}
T=-\rho+p_r+(n-2)p_t\;
\end{align}
is the trace of the energy momentum tensor $T^\mu{}_{\nu}$. 

\section{Upper bound on the photon sphere radius}\label{section3}
We now analyze the photon sphere and derive an upper bound for its radius. Due to the spherical symmetry of the system, we restrict our attention to particles moving in the equatorial plane, i.e., all polar angles satisfy
\begin{align}
	\theta_1=\theta_2=\cdots=\theta_{n-3}=\frac{\pi}{2} \;,\quad\theta_{n-2}=\phi \;,
\end{align}
where $\phi$ is the azimuthal coordinate. The Lagrangian for null geodesics in the spacetime (\ref{metrics}) is
\begin{align}
	\label{Lagrangian}
	2\mathcal{L}=-e^{-2\delta(r)}\mu(r)\dot{t}^2+\frac{\dot{r}^2}{\mu(r)}+r^2\dot{\phi}^2=0\;.
\end{align}
where a dot denotes differentiation with respect to an affine parameter. Since the Lagrangian is independent of $t$ and $\phi$, there are two conserved quantities: the energy $E$ and the angular momentum $\mathcal{L}$. From the Lagrangian (\ref{Lagrangian}), one derives the generalized momenta
\begin{align}
	\label{E}
	&p_t=-e^{-2\delta(r)}\mu(r)\dot{t}=-E\;,\\
	\label{L}
	&p_{\phi}=r^2\dot{\phi}=L\;,\\
	&p_r=\mu(r)^{-1}\dot{r}\;.
\end{align}
Substituting Eqs.~(\ref{E}) and (\ref{L}) into Eq.~(\ref{Lagrangian}) yields
\begin{align}
	\label{r1}
	\dot{r}^2=\mu\bigg(\frac{E^2}{e^{-2\delta}\mu}-\frac{L^2}{r^2}\bigg)
\end{align}
for the photon orbit. The corresponding effective potential can therefore be defined as
\begin{align}
	\label{V}
	V(r)=\mu\bigg(\frac{E^2}{e^{-2\delta}\mu}-\frac{L^2}{r^2}\bigg)\;.
\end{align}
A photon sphere must satisfy the two conditions $V(r)=0$ and $V'(r)=0$, which give
\begin{align}
	\label{PL}
	-r\mu'+2\mu(1+r\delta')=0\;.
\end{align}
Substituting Eqs.~(\ref{mu1}) and (\ref{delta1}) into Eq.~(\ref{PL}) leads to the characteristic equation
\begin{align}
	\label{cr}
	\mathcal{R}(r_\gamma)=0\;,
\end{align}
for the photon sphere, where 
\begin{align}
\label{R}
\mathcal{R}(r)=3-n+\mu(n-1)-\frac{16\pi r^2p_r}{n-2}\;.
\end{align}
From Eq.~(\ref{rhorH}), we obtain
\begin{align}
\label{prH}
-p_r(r_H)\leq \frac{(n-3)(n-2)}{16\pi r_H^2}\;.
\end{align}
We first prove the existence of a photon sphere exterior to the horizon.  At the horizon, using Eqs.~(\ref{rH}), (\ref{R}) and (\ref{prH}), we find
\begin{align}
\label{rh}
\mathcal{R}(r_H)\leq0\;.
\end{align}
Moreover, as $r\to\infty$, substituting Eqs.~(\ref{infinity}) into Eq.~(\ref{R}) together with the finite‑mass condition Eq.~(\ref{rho}) gives
\begin{align}
\label{Rinfty}
\mathcal{R}(r\rightarrow \infty)=2>0\;.
\end{align}
It is worth emphasizing that $\mathcal{R}(r)$ is a continuous function. Therefore, from Eqs.~(\ref{rh}) and (\ref{Rinfty}) we conclude that a photon sphere must exist in the region $r_H \le r < \infty$. This implies that there is no photon sphere in the region $r_H \le r < r_{\gamma}$ (where $r_\gamma$ denotes the radius of the innermost photon sphere), i.e.,
\begin{align}
	\label{unPS}
	\mathcal{R}(r_H \leq r < r_\gamma)<0\;.
\end{align}
To derive an upper bound, we follow the approach of Ref.~\cite{Hod:2013jhd} and introduce the pressure function $\mathcal{P}(r)=r^np_r(r)$. This construction, inspired by the four-dimensional case, will allow us to analyze the monotonicity of the radial pressure in a convenient form. Next, we shall deduce the behavior of $p_r$ by analyzing the properties of $\mathcal{P}(r)$. Differentiating $\mathcal{P}(r)$ with respect to $r$, we obtain
\begin{align}
\label{P'}
\mathcal{P}'(r)=\frac{r^{n-1}}{2\mu}[\mathcal{R}(\rho+p_r)+2\mu T]\;.
\end{align}
We assume that the matter field outside the black hole event horizon satisfies the following conditions:
\begin{itemize}
	\item [(1)] Weak Energy Condition (WEC): The energy density of the matter fields is positive semidefinite, and that it bounds the pressures
	\begin{align}
		\label{weak}
		\rho\geq 0\quad\;;\quad \rho\ge |p_r|,|p_t|\;.
	\end{align}
	\item [(2)]Non-positive trace condition: The trace of the energy-momentum tensor is assumed to be non-positive, this implies, i.e.,
	\begin{align}
		\label{Tle0}
		T\leq 0\;.
	\end{align}
	This condition holds for many common fields, including electromagnetic fields and conformally invariant matter, and is a natural extension of the assumption used in the four-dimensional proof~\cite{Hod:2013jhd}.
\end{itemize}
From Eq.~(\ref{rhorH}), one finds
\begin{align}
	\label{PrH}
	\mathcal{P}(r_H)=r^np_r(r_H)=-r^n\rho(r_H)\;.
\end{align}
Combining Eqs.~(\ref{weak}) and (\ref{PrH}), we find that the pressure function $\mathcal{P}(r)$ satisfies 
\begin{align}
	\label{PrH2}
	\mathcal{P}(r_H) \leq 0\;.
\end{align}
at the event horizon.

Next, we analyze the behavior of $\mathcal{P}(r)$ in the region from the event horizon to the photon sphere. By substituting the photon sphere characteristic inequality (\ref{unPS}), together with the energy conditions (\ref{weak}) and (\ref{Tle0}) for the matter field, into the pressure gradient (\ref{P'}), we obtain
\begin{align}
\label{unPSP'}
\mathcal{P}'(r_H\leq r<r_\gamma)<0\;.
\end{align}
This implies that the pressure function $\mathcal{P}(r)$ is monotonically decreasing in the region $r_H\leq r<r_\gamma$. Then, combining Eqs.~(\ref{PrH2}) and (\ref{unPSP'}), we finds that $\mathcal{P}(r)$ is a non-positive and monotonically decreasing function in the region $r_H\leq r<r_\gamma$, i.e.,
\begin{align}
\mathcal{P}(r)<0\quad\mathrm{where}\quad r_H\leq r<r_\gamma\;.
\end{align}
Accordingly, at the photon sphere, we have
\begin{align}
\label{prgamma}
p_r(r_\gamma)\leq 0\;.
\end{align}
Substituting Eq.~(\ref{prgamma}) into Eqs. (\ref{cr}) and (\ref{R}), we get
\begin{align}
\label{frgamma}
\mu(r_\gamma)\leq \frac{n-3}{n-1}\;.
\end{align}
From Eqs. (\ref{mu}) and (\ref{frgamma}), this yields the upper bound 
\begin{align}
\label{upper}
r_\gamma\leq [(n-1)m(r_\gamma)]^{\frac{1}{n-3}}
\end{align}
for the photon sphere. Since $m(r)$ is a non-decreasing function of $r$ and $m(r_\gamma)\le M$, where $M=m(r\to\infty)$ is the total ADM mass of the black hole spacetime, we finally obtain the upper bound
\begin{align}
\label{upper bound}
r_\gamma\leq [(n-1)M]^{\frac{1}{n-3}}\;,
\end{align}
In the four-dimensional case, this reduces to $r_\gamma\leq 3M$, which is consistent with the result in Ref.~\cite{Hod:2013jhd,Yang:2019zcn}. It is worth noting that this upper bound $3M$ is saturated by the photon sphere of the vacuum (bald) Schwarzschild black hole. Therefore, it is reasonable to hypothesize that the upper bound of the photon sphere radius is saturated by a vacuum (bald) black hole, and the presence of matter fields alters this radius, making it smaller than that of the hairless black hole.

\section{Lower bound on the photon sphere radius}\label{section4}
To obtain the lower bound, we need an additional condition:
\begin{itemize}
	\item [(3)] Monotonicity of the radial pressure function: The radial pressure function
	\begin{align}
		\label{P}
		P(r)\equiv |r^{n-1}p_r(r)|
	\end{align}
	decreases monotonically with $r$. This condition generalizes the four-dimensional assumption used by~\cite{Hod:2020pim} and is motivated by the behavior of matter fields in many known hairy black hole solutions, where the radial pressure typically decays rapidly outside the horizon.
\end{itemize}
Following Ref.~\cite{Hod:2020pim}, we first derive a lower bound on the mass of the matter fields exterior to the horizon in higher-dimensional black holes. From Eq.~(\ref{m}) and condition (\ref{weak}), one obtains
\begin{align}
	\label{mhair}
	m_{\mathrm{ex}}=\frac{8\pi}{n-2}\int_{r_H}^{\infty}r^{n-2}\rho(r)dr\ge -\frac{8\pi}{n-2}\int_{r_H}^{\infty}r^{n-2}p_r(r)dr\;.
\end{align}
Crucially, as shown in Sec.~\ref{section3}, it has been shown that under the conditions (\ref{weak}) and (\ref{Tle0}), the function $\mathcal{P}(r)\equiv r^n p_r$ is non-positive and monotonically decreasing in the interval $r\in [r_H,r_\gamma]$:
\begin{align}
	\label{decreasing}
	\{p_r(r)\le 0\quad\mathrm{and}\quad (r^np_r)'\le 0\}\quad \mathrm{for} \quad r_H\le r\le r_\gamma\;.
\end{align}
where $r_\gamma$ is the radius of photon sphere. Consequently,
\begin{align}
	\label{relation}
	0\le -r^n_Hp_r(r_H)\le -r^np_r(r)\quad \mathrm{for}\quad r_H\le r\le r_\gamma\;.
\end{align}
Using the relation (\ref{relation}) in Eq.~(\ref{mhair}) gives the chain of inequalities
\begin{align}
	\label{mhairinequalities}
	m_{\mathrm{ex}}\ge -\frac{8\pi}{n-2}\int_{r_H}^{r_\gamma}r^{n-2}p_r(r)dr\ge -\frac{8\pi}{n-2}\int_{r_H}^{r_\gamma}\frac{r_H^{n}p_r(r_H)}{r^2}dr=-\frac{8\pi}{n-2}r^n_Hp_r(r_H)\bigg(\frac{1}{r_H}-\frac{1}{r_\gamma}\bigg)
\end{align}
for the mass $m_{\mathrm{ex}}$ of the external matter fields.

The location of the photon sphere is determined by the condition (\ref{cr})
\begin{align}
	\label{R=0}
	\mathcal{R}(r_\gamma)=3-n+\mu(n-1)-\frac{16\pi r_\gamma^2p_r(r_\gamma)}{n-2}=0\;.
\end{align}
Substituting the lower bound (\ref{mhairinequalities}) into Eq.~(\ref{R=0}) and using the relation (\ref{mu}), one obtains the inequality
\begin{align}
	\label{inequality1}
	2-(n-1)\frac{r^{n-3}_H}{r^{n-3}_\gamma}-\frac{16\pi r^2_\gamma p_r(r_\gamma)}{n-2}+\frac{16\pi(n-1)}{(n-2)r^{n-3}_\gamma}r^n_Hp_r(r_H)\bigg(\frac{1}{r_H}-\frac{1}{r_\gamma}\bigg)\ge 0\;,
\end{align}
which characterizes the null circular geodesics of the static spherically symmetric black hole spacetime. In addition, using the inequality $r^{n-1}_Hp_r(r_H)\le r^{n-1}_\gamma p_r(r_\gamma)$, which follows from the assumed monotonic behavior of the radial pressure function (\ref{P}) and the fact that the radial pressure is non-positive between the black hole horizon and the innermost null circular geodesic, one deduces
from (\ref{inequality1}) the inequality
\begin{align}
	\label{inequality2}
	2-(n-1)\frac{r^{n-3}_H}{r^{n-3}_\gamma}-\frac{16\pi r^2_\gamma p_r(r_\gamma)}{n-2}+\frac{16\pi(n-1)}{(n-2)r^{n-3}_\gamma}r_Hr_\gamma^{n-1}p_r(r_\gamma)\bigg(\frac{1}{r_H}-\frac{1}{r_\gamma}\bigg)\ge 0\;,
\end{align}
Simplifying the above equation, we have
\begin{align}
	\label{inequality3}
	2-(n-1)\frac{r^{n-3}_H}{r^{n-3}_\gamma}+\frac{16\pi r^2_\gamma p_r(r_\gamma)}{n-2}\bigg[n-2-(n-1)\frac{r_H}{r_\gamma}\bigg]\ge 0\;.
\end{align}
Substituting the condition (\ref{R=0}) into Eq.~(\ref{inequality3}), we have
\begin{align}
	\label{inequality4}
	2-(n-1)\frac{r^{n-3}_H}{r^{n-3}_\gamma}+[(n-1)\mu(r_\gamma)-(n-3)]\bigg[n-2-(n-1)\frac{r_H}{r_\gamma}\bigg]\ge 0\;.
\end{align}
Introducing $x=\frac{r_\gamma}{r_H}>1$, so $r^{n-3}_\gamma/r^{n-3}_H=x^{n-3}$, and let $\mu_\gamma=\mu(r_\gamma)$. Then Eq.~(\ref{inequality4}) becomes
\begin{align}
	\label{inequality5}
	2-(n-1)x^{-(n-3)}+[(n-1)\mu_\gamma-(n-3)]\bigg[n-2-(n-1)x^{-1}\bigg]\ge 0\;.
\end{align}
Define
\begin{align}
	\label{G}
	G(x)=2-(n-1)x^{-(n-3)}+[(n-1)\mu_\gamma-(n-3)]\bigg[n-2-(n-1)x^{-1}\bigg]\;,
\end{align}
so that $G(x)\ge 0$. Since $p_r(r_\gamma)\le 0$, Eq.~(\ref{R=0}) implies 
\begin{align}
	A\equiv (n-1)\mu_\gamma-(n-3)\le 0\;.
\end{align}
We now analyse the function $G(x)$ for $x>1$. Define
\begin{align}
	f(x)=2-(n-1)x^{-(n-3)}\;,\quad g(x)\equiv n-2-(n-1)x^{-1}\;,
\end{align}
so that $G(x)=f(x)+Ag(x)$ with $A\le 0$\;.

Consider two cases according to the sign of $g(x)$.
\begin{itemize}
	\item[(1).] $g(x)\ge 0$ (i.e., $x\ge\frac{n-1}{n-2}$): Then $Ag(x)\le 0$ and consequently $G(x)\le f(x)$. The condition $G(x)\ge 0$ therefore requires $f(x)\ge 0$, i.e.,
	\begin{align}
		2-(n-1)x^{-(n-3)}\ge 0\;,
	\end{align}
	Hence, in this case we must have
	\begin{align}
		x\ge \bigg(\frac{n-1}{2}\bigg)^{\frac{1}{n-3}}\;.
	\end{align}
	
	\item[(2).] $g(x)<0$ (i.e. $1<x<\frac{n-1}{n-2}$): we prove that in this case, one actually always has $G(x)< 0$. This is done by observing that
	\begin{align}
		\frac{\partial G}{\partial A}=g(x)<0\;,
	\end{align}
	hence $G(x)$ attains its maximum at the smallest allowed value of $A$. The smallest possible $A$ is $A_{\mathrm{min}}=-(n-3)$ (corresponding to $\mu_r=0$). Substituting $A_{\mathrm{min}}=-(n-3)$ into Eq.~(\ref{G}) gives
	\begin{align}
		G_{\mathrm{max}}(x)=(n-1)[-(n-4)-x^{-(n-3)}+(n-3)x^{-1}]\;.
	\end{align}
	For $n\ge 5$, $G_{\mathrm{max}}(x)$ is strictly negative for all $x>1$  (it vanishes at $x=1$ and is decreasing\footnote{From
		\begin{align}
			G_{\mathrm{max}}(x)=(n-1)[-(n-4)-x^{-(n-3)}+(n-3)x^{-1}]\;,
		\end{align}
		we have $G_{\mathrm{max}}(1)=0$, and
		\begin{align}
			G'_{\mathrm{max}}(x)=(n-1)[(n-3)x^{-(n-2)}-(n-3)x^{-2}]=(n-1)(n-3)x^{-2}\bigg(x^{-(n-4)}-1\bigg)\;.
		\end{align}
		For $n\ge 5$, we have $n-4\ge 1$ and $x^{-(n-4)}<1$ for $x>1$, so $G'_{\mathrm{max}}(x)<0$.}). For $n=4$,  from Eq.~(\ref{G}) one obtains $G(x)=3\mu_\gamma(2-3x^{-1})$, which is negative precisely when $1<x<\frac{3}{2}$. Thus, for any $n\ge 4$, if $g(x)<0$, then $G(x)<0$, which conflicts with Eq.~(\ref{inequality5}).
\end{itemize}
Combining the two cases we conclude that from the condition $G(x)\ge 0$, we have
\begin{align}
	x\ge\frac{n-1}{n-2}\;,\quad\mathrm{and}\quad x\ge \bigg(\frac{n-1}{2}\bigg)^{\frac{1}{n-3}}\;.
\end{align}
Using the result (\ref{A1}) in appendix~\ref{Appendix},  we obtain the necessary and sufficient condition for $G(x)\ge 0$:
\begin{align}
	x\ge \bigg(\frac{n-1}{2}\bigg)^{\frac{1}{n-3}}\;.
\end{align}
Returning to the original variables, this inequality becomes
\begin{align}
	\label{lower_bound}
	r_\gamma\ge \bigg(\frac{n-1}{2}\bigg)^{\frac{1}{n-3}}r_H\;,
\end{align}
which provides a lower bound on the photon-sphere radius in higher-dimensional black holes.

\section{Discussion and conclusion}\label{conclusion}
In this work, we derived both an upper and a lower bound for the photon sphere radius of static, spherically symmetric, asymptotically flat black holes in $n$-dimensional Einstein gravity ($n\ge 4$), under physically motivated assumptions on the matter fields. The upper bound,
\begin{align}
r_\gamma\le [(n-1)M]^{\frac{1}{n-3}}\;,
\end{align}
generalises previous four‑dimensional result~\cite{Hod:2013jhd,Yang:2019zcn} and is saturated by the vacuum Tangherlini black hole. The lower bound,
\begin{align}
r_\gamma\ge (\frac{n-1}{2})^{1/(n-3)}r_H
\end{align}
requires the additional condition that $|r^{n-1}p_r(r)|$ decreases monotonically; it reduces to $r_\gamma\ge \frac{3}{2}r_H$ for $n=4$, in agreement with~\cite{Hod:2020pim,Yang:2019zcn}. Together, these bounds provide a complete characterisation of the possible location of the photon sphere in this class of spacetimes.

It is important to emphasize the scope and limitations of our result. Our derivation relies on the Tangherlini-type metric ansatz (\ref{metrics}), which assumes a direct product structure with a maximally symmetric transverse space and strict asymptotic flatness in all $n$ dimensions. This framework, while natural for a first generalization, excludes more intricate geometries such as warped spacetimes, brane-world scenarios, and models with compactified extra dimensions. In such settings, the properties of the photon sphere may differ substantially, and our bound may not apply.

Furthermore, we must address the potential observational implications---or lack thereof---for astrophysical black holes. In phenomenologically viable extra-dimensional models, the additional spatial dimensions are typically compactified at a scale far below the micrometer level, as constrained by current laboratory and astrophysical experiments. For solar-mass or supermassive black holes, whose horizons are on the order of kilometers to astronomical units, the effective gravitational dynamics at macroscopic scales reduces to four dimensions. Consequently, any corrections to the photon sphere radius arising from the presence of such small extra dimensions would be overwhelmingly negligible. Therefore, our result does not translate to observable deviations from the four-dimensional bound for astrophysical black holes, unless one entertains scenarios with non-compact or extremely large extra dimensions, which are already strongly disfavored by existing constraints.

Notwithstanding these limitations, our work provides a valuable theoretical extension within a well-defined class of higher-dimensional black holes. The bound offers a geometric constraint that may inform the study of black hole shadows in idealized higher-dimensional contexts and contributes to the toolkit for analyzing black holes in theories beyond standard four-dimensional general relativity.

Future research could extend this work in several promising directions. Possible extensions include investigating axisymmetric (rotating) configurations~\cite{Cunha:2017qtt,Cunha:2020azh}, considering asymptotically de Sitter or anti-de Sitter spacetimes, or examining the implications of quantum corrections on the energy conditions. Such generalizations would further illuminate the interplay among energy conditions, horizon geometry, and null geodesics in curved spacetimes.
	
	
\appendix
\section{Proof of the inequality $\frac{n-1}{n-2}\le (\frac{n-1}{2})^{1/(n-3)}$ for $n\ge 4$}\label{Appendix}
We prove that for $n\ge 4$,
\begin{align}
	\label{A1}
	\frac{n-1}{n-2}\le \bigg(\frac{n-1}{2}\bigg)^{\frac{1}{n-3}}
\end{align}
with equality if and only if $n=4$. 

Set $x=n-1(\ge 3)$. Then the inequality becomes
\begin{align}
	\frac{x}{x-1}\le\bigg(\frac{x}{2}\bigg)^{\frac{1}{x-2}}\;.
\end{align}
Taking natural logarithms on both sides, it is equivalent to
\begin{align}
	\ln\bigg(\frac{x}{x-1}\bigg)\le \frac{1}{x-2}\ln\bigg(\frac{x}{2}\bigg)\;,
\end{align}
or, rearranging,
\begin{align}
	(x-2)\ln\bigg(\frac{x}{x-1}\bigg)-\ln\bigg(\frac{x}{2}\bigg)\le 0\;.
\end{align}
Define
\begin{align}
	F(x)\equiv (x-2)\ln\bigg(\frac{x}{x-1}\bigg)-\ln\bigg(\frac{x}{2}\bigg),\quad x\ge 3\;,
\end{align}
we have $F(3)=0$. Its derivative is
\begin{align}
	F'(x)=\ln\bigg(\frac{x}{x-1}\bigg)-\frac{2x-3}{x(x-1)}\;.
\end{align}
Using the elementary inequality $\ln (1+t)<t$ for $t>0$ with $t=\frac{1}{x-1}$, we obtain
\begin{align}
	\ln\bigg(\frac{x}{x-1}\bigg)=\ln\bigg(1+\frac{1}{x-1}\bigg)<\frac{1}{x-1}\;.
\end{align}
For $x>3$,  it is easy to verify that
\begin{align}
	\frac{1}{x-1}\le \frac{2x-3}{x(x-1)}\;.
\end{align}
Hence, for all $x>3$,
\begin{align}
	\ln\bigg(\frac{x}{x-1}\bigg)<\frac{1}{x-1}<\frac{2x-3}{x(x-1)}\;,
\end{align}
which implies $F'(x)<0$ for $x>3$. Consequently, $F(x)$ is strictly decreasing on $[3,\infty)$ and therefore $F(x)<F(3)=0$ for every $x>3$. This proves (\ref{A1}) with strict inequality for $x>3$ (i.e., for $n>4$), while equality holds at $x=3$ (i.e., at $n=4$).



\begin{thebibliography}{99}
\bibitem{EventHorizonTelescope:2019dse}
K.~Akiyama \textit{et al.} [Event Horizon Telescope],
Astrophys. J. Lett. \textbf{875}, L1 (2019)
doi:10.3847/2041-8213/ab0ec7
[arXiv:1906.11238 [astro-ph.GA]].

\bibitem{EventHorizonTelescope:2022wkp} 
K.~Akiyama \textit{et al.} [Event Horizon Telescope], 
Astrophys. J. Lett. \textbf{930}, no.2, L12 (2022) 
doi:10.3847/2041-8213/ac6674 
[arXiv:2311.08680 [astro-ph.HE]].

\bibitem{Perlick:2010zh} 
V.~Perlick, 
Living Rev. Relativ. 7, 9 (2004)
doi.org/10.12942/lrr-2004-9
[arXiv:1010.3416 [gr-qc]].


\bibitem{Falcke:1999pj} 
H.~Falcke, F.~Melia and E.~Agol, 
Astrophys. J. Lett. \textbf{528}, L13 (2000) 
doi:10.1086/312423 
[arXiv:astro-ph/9912263 [astro-ph]].
	
\bibitem{Cardoso:2008bp}
V.~Cardoso, A.~S.~Miranda, E.~Berti, H.~Witek and V.~T.~Zanchin,
Phys. Rev. D \textbf{79}, no.6, 064016 (2009)
doi:10.1103/PhysRevD.79.064016
[arXiv:0812.1806 [hep-th]].
		
		
\bibitem{Berti:2009kk}
E.~Berti, V.~Cardoso and A.~O.~Starinets,
Class. Quant. Grav. \textbf{26}, 163001 (2009)
doi:10.1088/0264-9381/26/16/163001
[arXiv:0905.2975 [gr-qc]].

\bibitem{Nunez:1996xv} 
D.~Nunez, H.~Quevedo and D.~Sudarsky, 
Phys. Rev. Lett. \textbf{76}, 571-574 (1996) 
doi:10.1103/PhysRevLett.76.571 
[arXiv:gr-qc/9601020 [gr-qc]]. 


\bibitem{Hod:2011aa}
S.~Hod,
Phys. Rev. D \textbf{84}, 124030 (2011)
doi:10.1103/PhysRevD.84.124030
[arXiv:1112.3286 [gr-qc]].


\bibitem{Hod:2013jhd}
S.~Hod,
Phys. Lett. B \textbf{727}, 345-348 (2013)
doi:10.1016/j.physletb.2013.10.047
[arXiv:1701.06587 [gr-qc]].

\bibitem{Yang:2019zcn}
R.~Q.~Yang and H.~Lu,
Eur. Phys. J. C \textbf{80} (2020) no.10, 949
doi:10.1140/epjc/s10052-020-08521-7
[arXiv:2001.00027 [gr-qc]].



\bibitem{Hod:2020pim}
S.~Hod,
Phys. Rev. D \textbf{101}, no.8, 084033 (2020)
doi:10.1103/PhysRevD.101.084033
[arXiv:2012.03962 [gr-qc]].

\bibitem{Lu:2019zxb}
H.~Lu and H.~D.~Lyu,
Phys. Rev. D \textbf{101} (2020) no.4, 044059
doi:10.1103/PhysRevD.101.044059
[arXiv:1911.02019 [gr-qc]].

\bibitem{Peng:2020shc} 
Y.~Peng, 
Eur. Phys. J. C \textbf{80}, no.8, 755 (2020) 
doi:10.1140/epjc/s10052-020-8358-z 
[arXiv:2006.02618 [gr-qc]].

\bibitem{Hod:2023jmx}
S.~Hod,
JHEP \textbf{12}, 178 (2023)
doi:10.1007/JHEP12(2023)178
[arXiv:2311.17462 [gr-qc]].

\bibitem{Liu:2024odv}
G.~Liu and Y.~Peng,
Eur. Phys. J. C \textbf{84}, no.7, 685 (2024)
doi:10.1140/epjc/s10052-024-13053-5
[arXiv:2402.15517 [gr-qc]].

\bibitem{Song:2025fgq} 
Y.~Song, J.~Fu and Y.~Cen, 
Eur. Phys. J. C \textbf{85}, no.9, 981 (2025) 
doi:10.1140/epjc/s10052-025-14727-4 
[arXiv:2508.19823 [gr-qc]]. 

\bibitem{Cao:2019vlu}
L.~M.~Cao and Y.~Song,
Eur. Phys. J. C \textbf{81}, no.8, 714 (2021)
doi:10.1140/epjc/s10052-021-09502-0
[arXiv:1910.13758 [gr-qc]].


\bibitem{Lee:2025xdm} 
H.~Lee and B.~Gwak, 
[arXiv:2511.13086 [gr-qc]].


\bibitem{Volkov:1998cc}
M.~S.~Volkov and D.~V.~Gal'tsov,
Phys. Rept. \textbf{319}, 1-83 (1999)
doi:10.1016/S0370-1573(99)00010-1
[arXiv:hep-th/9810070 [hep-th]].

\bibitem{Volkov:2016ehx}
M.~S.~Volkov,
doi:10.1142/9789813226609{\_}0184
[arXiv:1601.08230 [gr-qc]].


\bibitem{Cunha:2017qtt}
P.~V.~P.~Cunha, E.~Berti and C.~A.~R.~Herdeiro,
Phys. Rev. Lett. \textbf{119} (2017) no.25, 251102
doi:10.1103/PhysRevLett.119.251102
[arXiv:1708.04211 [gr-qc]].

\bibitem{Cunha:2020azh}
P.~V.~P.~Cunha and C.~A.~R.~Herdeiro,
Phys. Rev. Lett. \textbf{124} (2020) no.18, 181101
doi:10.1103/PhysRevLett.124.181101
[arXiv:2003.06445 [gr-qc]].


\end{thebibliography}
\end{document}